\documentclass[12pt]{article}
\usepackage{amsfonts}
\usepackage{amsmath}
\usepackage{amssymb}
\usepackage{amsthm}
\usepackage{geometry}
\usepackage{rotating}
\usepackage{color}

\usepackage[utf8]{inputenc}
\usepackage{array}
\usepackage[english]{babel}
\usepackage{float}

\usepackage{bm}
\usepackage{natbib}

\newcommand{\eqd}   {\buildrel d \over =}

\begin{document}
\title{On split sample and randomized confidence intervals for binomial proportions}

\author
{ {M{\aa}ns Thulin}\\{\footnotesize{Department of Mathematics, Uppsala University}}}
\date{}

\maketitle


\begin{abstract}
\noindent 
Split sample methods have recently been put forward as a way to reduce the coverage oscillations that haunt confidence intervals for parameters of lattice distributions, such as the binomial and Poisson distributions. We study split sample intervals in the binomial setting, showing that these intervals can be viewed as being based on adding discrete random noise to the data. It is shown that they can be improved upon by using noise with a continuous distribution instead, regardless of whether the randomization is determined by the data or an external source of randomness. We compare split sample intervals to the randomized Stevens interval, which removes the coverage oscillations completely, and find the latter interval to have several advantages.
\noindent 
   \\[1.5mm] {\bf Keywords:} Binomial distribution; confidence interval; Poisson distribution; randomization; split sample method.
\end{abstract}

\section{Introduction}\label{introduction}
The problem of constructing confidence intervals for parameters of discrete distributions continues to attract considerable interest in the statistical community. The lack of smoothness of these distributions causes the coverage (i.e. the probability that the interval covers the true parameter value) of such intervals to fluctuate from $1-\alpha$ when either the parameter value or the sample size $n$ is altered. Recent contributions to the theory of these confidence intervals include \citet{peng11}, \citet{newcombe11}, \citet{go1}, \citet{lurz} and \citet{thu2}.

The purpose of this short note is to discuss the split sample method recently proposed by \citet{hall14}. The split sample method reduces the oscillations of the coverage by splitting the sample in two. It is applicable to most existing confidence intervals for parameters of lattice distributions, including the binomial and Poisson distributions. For simplicity, we will in most of the remainder of the paper limit our discussion to the binomial setting, with the split sample method being applied to the celebrated \citet{wilson27} interval for the proportion $p$. Our conclusions are however equally valid for other confidence intervals and distributions.

A random variable $X\sim \mbox{Bin}(n,p)$ is the sum of a sequence $X_1,\ldots,X_n$ of independent $\mbox{Bernoulli}(p)$ random variables. The split sample method is applied to $X_1,\ldots,X_n$ rather than to $X$. The idea is to split the sample into two sequences $X_1,\ldots,X_{n_1}$ and $X_{n_1+1},\ldots,X_{n_1+n_2}$ with $n_1+n_2=n$ and $n_1\neq n_2$. In the formula for the chosen confidence interval, the maximum likelihood estimator $\hat{p}=X/n$ is then replaced by the weighted estimator
\begin{equation}\label{eq1}
\tilde{p}=\frac{1}{2}\Big(\frac{1}{n_1}\sum_{i=1}^{n_1}X_i+\frac{1}{n_2}\sum_{i=n_1+1}^{n_1+n_2}X_i\Big).
\end{equation}
The formula for the Wilson interval is
\[
\frac{\hat{p}n+z_{\alpha/2}^2/2}{n+z_{\alpha/2}^2}\pm \frac{z_{\alpha/2}}{n+z_{\alpha/2}^2}\sqrt{\hat{p}(1-\hat{p})n+z_{\alpha/2}^2/4},
\]
where $z_{\alpha/2}$ is the $\alpha/2$ standard normal quantile, so the split sample Wilson interval is given by
\[
\frac{\tilde{p}n+z_{\alpha/2}^2/2}{n+z_{\alpha/2}^2}\pm \frac{z_{\alpha/2}}{n+z_{\alpha/2}^2}\sqrt{\tilde{p}(1-\tilde{p})n+z_{\alpha/2}^2/4}.
\]

\citet{hall14} showed by numerical and asymptotic arguments based on \citet{hall13} that when $n_1=n/2+0.15n^{3/4}$ (rounded to the nearest integer) and $n_2=n-n_1$ the split sample method greatly reduces the oscillations of the coverage of the interval, without increasing the interval length. In the remainder of the paper, whenever $n_1$ and $n_2$ need to be specified, we will use these values.

Depending on how the sample is split, different confidence intervals will result, making the split sample interval a randomized confidence interval. If the sequence $X_1,\ldots,X_n$ is available, the interval can be data-randomized in the sense that the randomization can be determined by the data: the first $n_1$ observations can be put in the first subsample and the remaining $n_2$ observation in the second subsample. If the results of the individual trials have not been recorded, one must  use randomness from outside the data to create a sequence $X_1,\ldots,X_n$ of 0's and 1's such that $\sum_{i=1}^nX_i=X$. We will discuss these two settings separately.

\citet{hall14} left two questions open. The first is how the split sample interval performs in comparison to other randomized intervals, as \citet{hall14} only compared the split sample interval to non-randomized confidence intervals. The second question is to what extent the randomization can affect the bounds of the interval. In the remainder of this note we answer these questions, discussing the impact of the random splitting on the confidence interval and comparing the split sample interval to alternative intervals. Section \ref{randomized} is concerned with externally randomized intervals whereas we in Section \ref{datarandomized} study data-randomized intervals. In both settings, we find the competing intervals to be superior to the split sample interval. Various criticisms of randomized intervals are then discussed in Section \ref{discussion}.

\section{Externally randomized confidence intervals}\label{randomized}

\subsection{Split sample and alternative intervals}
\begin{figure}
\begin{center}
 \caption{The distribution of $Y|X$ when $n=11$, $n_1=6$ and $n_2=5$.}\label{fig0}
   \includegraphics[width=\textwidth]{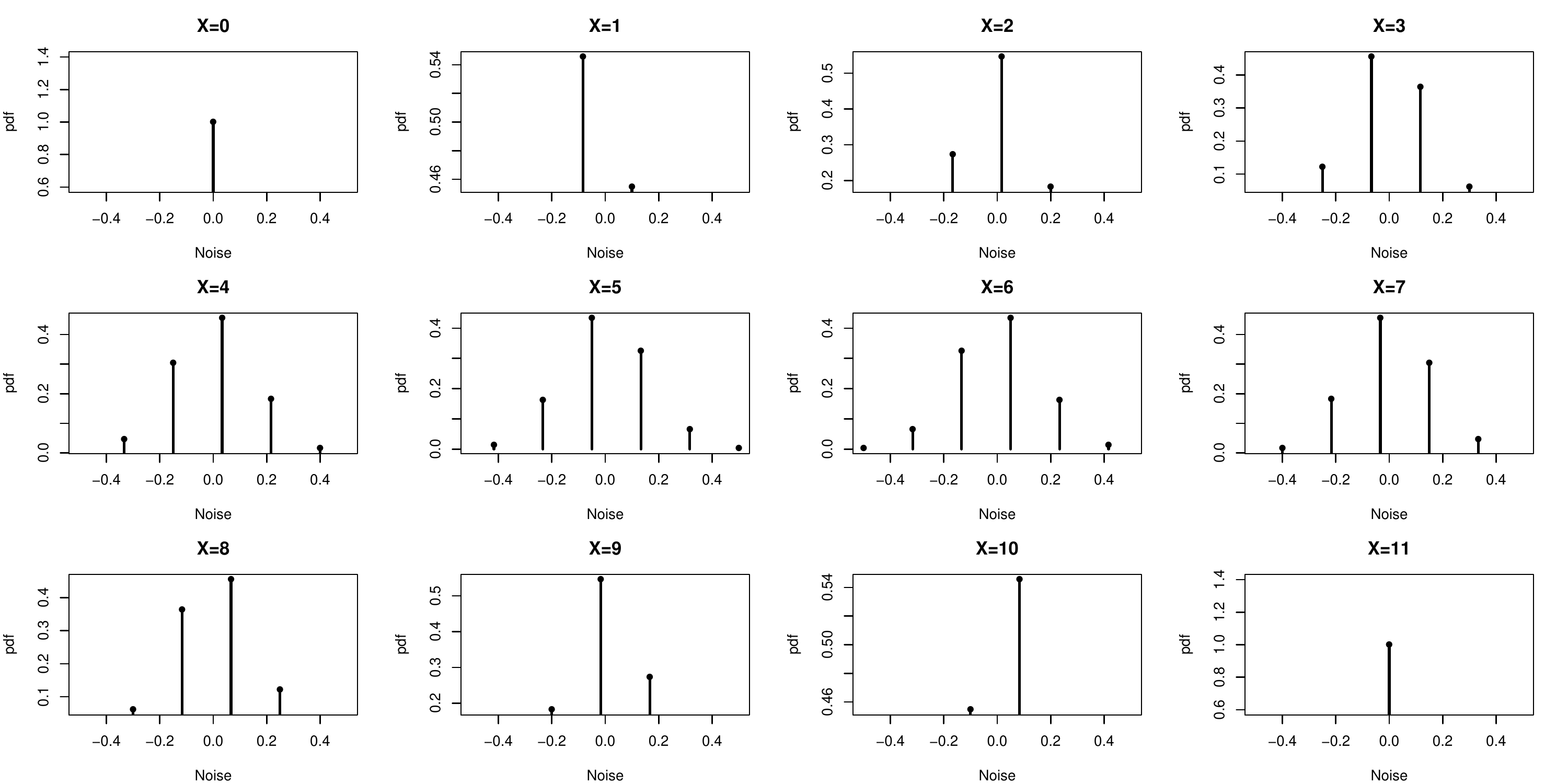}
 \end{center}
\end{figure}

A strategy for smoothing the distribution of the binomial random variable $X$ is to base our inference on $X+Y$, where $Y$ is a comparatively small random noise, using $X+Y$ instead of $X$ in the formula for our chosen confidence interval. Having a smoother distribution leads to a better normal approximation, which in turn reduces the coverage fluctuations of the interval. The split sample method can be seen to be a special case of this strategy. Let $Z$ be a random variable which, conditioned on $X$, follows a $\mbox{Hypergeometric}(n,X,n_1)$ distribution. Then it follows from (\ref{eq1}) that
$$\tilde{X}=n\tilde{p}\eqd\frac{n}{2n_1}Z-\frac{n}{2n_2}(X-Z),$$
so that
$$\tilde{X}\eqd X+Y\qquad\mbox{with}\qquad Y=\frac{n}{2n_1}Z-\frac{n}{2n_2}Z+\frac{n_1-n_2}{2n_2}X.$$
The conditional distribution of $Y$ when $n=11$ is shown in Figure \ref{fig0}. 
Indeed, when the sequence $X_1,\ldots,X_n$ has not been observed, the splitting of the sample is superficial (since the sample itself is not available to us) and it may be more natural to think of the split sample procedure as adding this discrete random noise term $Y$, the distribution of which is conditioned on $X$. 

\begin{figure}
\begin{center}
 \caption{Comparison between the cumulative distribution functions of normal distributions ({\color{red}red}) and (a) $X$, (b) $\tilde{X}$, (c) $\dot{X}$ when $X\sim \mbox{Bin}(7,1/2)$.}\label{fig1}
   \includegraphics[width=\textwidth]{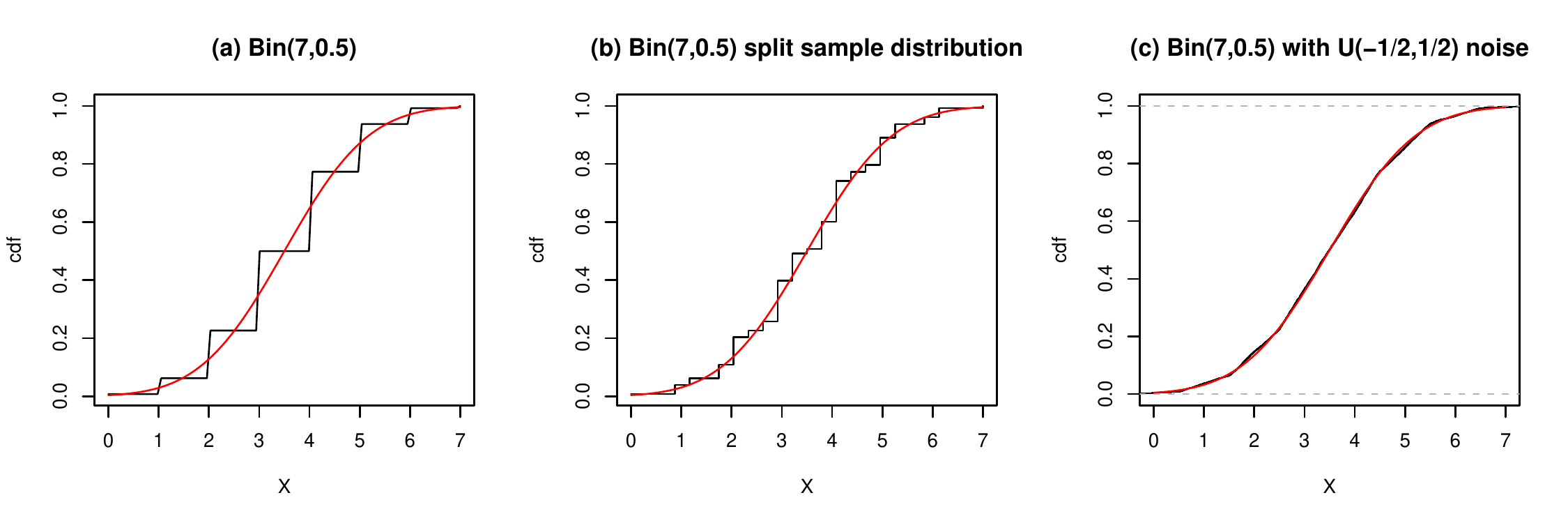}
 \end{center}
\end{figure}

If one has decided to use random noise to improve the normal approximation, the next step is to ask whether there are other distributions for the noise that yield an even better approximation and thereby decrease the coverage oscillations even further. The answer is yes, for instance if $X$ is replaced by $\dot{X}=X+Y$ where $Y\sim \mbox{U}(-1/2,1/2)$ independently of $X$. The normal approximations of $X$, $\tilde{X}$ and $\dot{X}$ are compared in Figure \ref{fig1}. We will refer to the interval based on $\dot{X}$ as the $\mbox{U}(-1/2,1/2)$ interval.

A randomized confidence interval that does not rely on a normal approximation is the \citet{stevens50} interval. It belongs to an important class of confidence intervals for $p$, consisting of intervals $(p_L,p_U)$ where the lower bound $p_L$ is such that
$$ \nu_1\cdot \binom{n}{X}p_L^X(1-p_L)^{n-X}+\sum_{k=X+1}^n\binom{n}{k}p_L^k(1-p_L)^{n-k}=\alpha/2$$
and the upper bound $p_L$ is such that
$$ \nu_2\cdot \binom{n}{X}p_U^X(1-p_U)^{n-X}+\sum_{k=0}^{X-1}\binom{n}{k}p_U^k(1-p_U)^{n-k}=\alpha/2.$$
For $\nu_1=\nu_2=1$ this is the non-randomized Clopper--Pearson interval \citep{cp1} and for $\nu_1=\nu_2=1/2$ this is the non-randomized mid-p interval \citep{lancaster61}. The choice $\nu_1\sim \mbox{U}(0,1)$ and $\nu_2=1-\nu_1$ corresponds to the randomized Stevens interval, which is an inversion of the most powerful unbiased test for $p$ \citep{blank56}. We note that Stevens' procedure also can be applied when constructing confidence intervals for parameters of Poisson and negative binomial distributions.

\subsection{Comparison of externally randomized intervals}\label{comp1}
\begin{figure}
\begin{center}
 \caption{Coverage and expected length of some confidence intervals for the binomial proportion $p$.}\label{fig2}
   \includegraphics[width=\textwidth]{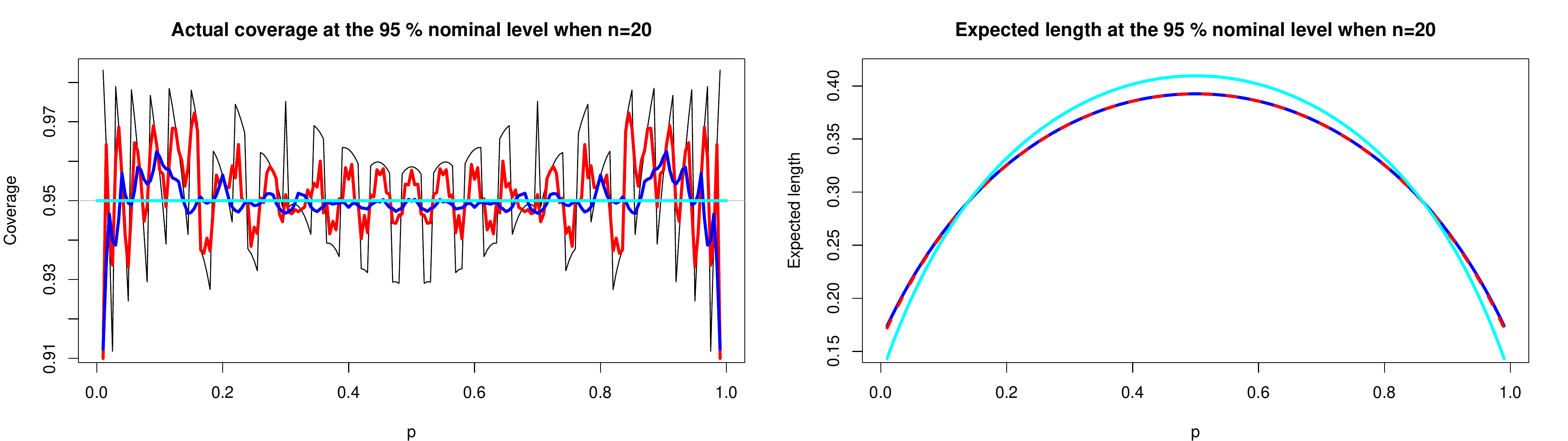}
 \includegraphics[width=\textwidth]{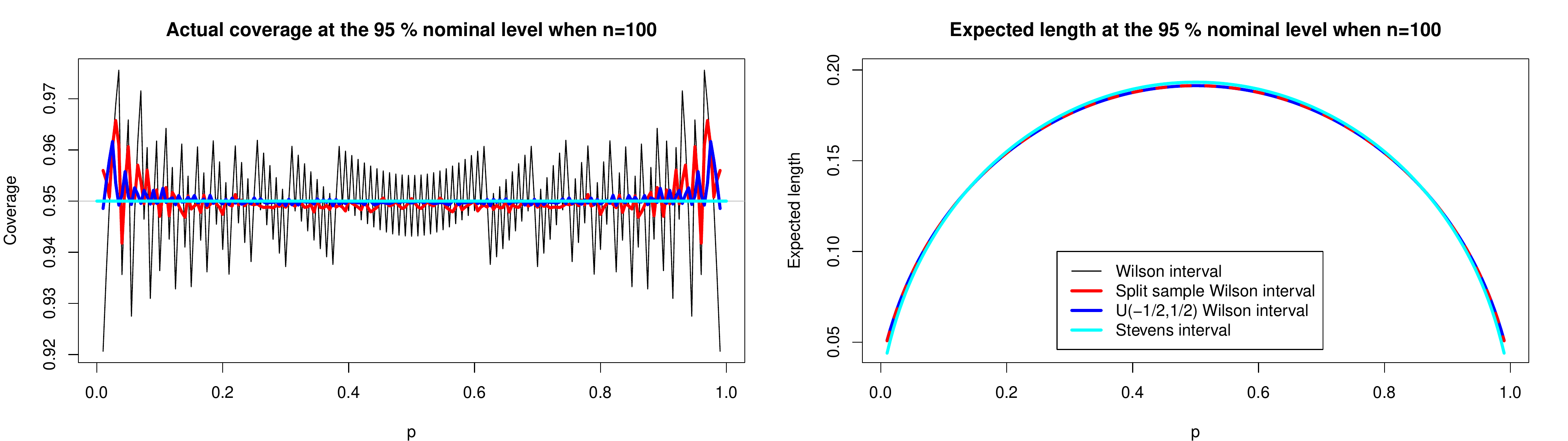}
 \end{center}
\end{figure}

The coverage fluctuations and expected length of the split sample Wilson, $\mbox{U}(-1/2,1/2)$ Wilson and Stevens intervals are compared in Figure \ref{fig2}. The $\mbox{U}(-1/2,1/2)$ interval improves upon the split sample interval about as much as the split sample interval improves upon the standard Wilson interval. Unlike the split sample and $\mbox{U}(-1/2,1/2)$ intervals, the Stevens interval has coverage exactly equal to $1-\alpha$ for all values of $p$, so that there are no coverage oscillations at all. The difference in expected length between the intervals is quite small for all combinations of $p$ and $n$ and is decreasing in $n$. The Stevens interval is shorter when $p$ is close to 0 or 1, but wider when $p$ is close to $1/2$. When $n=20$ the difference in expected length is at most 0.018 and when $n=100$ it is at most 0.002.


The randomization has a strong impact on the bounds of the confidence intervals. As an example, when $n=10$ and $X=2$ the upper bound of the split sample Wilson interval ranges from 0.476 to 0.558, meaning that the range of the upper bound is 0.082 or 18 \% of the interval's conditional expected length 0.45. The impact of the randomization can be substantial even for relatively large $n$. Consider for instance the case $n=200$, $X=40$. The conditional mean length of the split sample Wilson interval is 0.11, whereas the range of the upper bound of the interval is 0.016, or 14.5 \% of the conditional expected length of the interval.

\begin{figure}
\begin{center}
 \caption{Range of the upper bounds of randomized confidence intervals for different combinations of $n$ and $X$. The ranges are only defined for integer $X$; the lines are used only to guide the eye.}\label{fig3}
   \includegraphics[width=\textwidth]{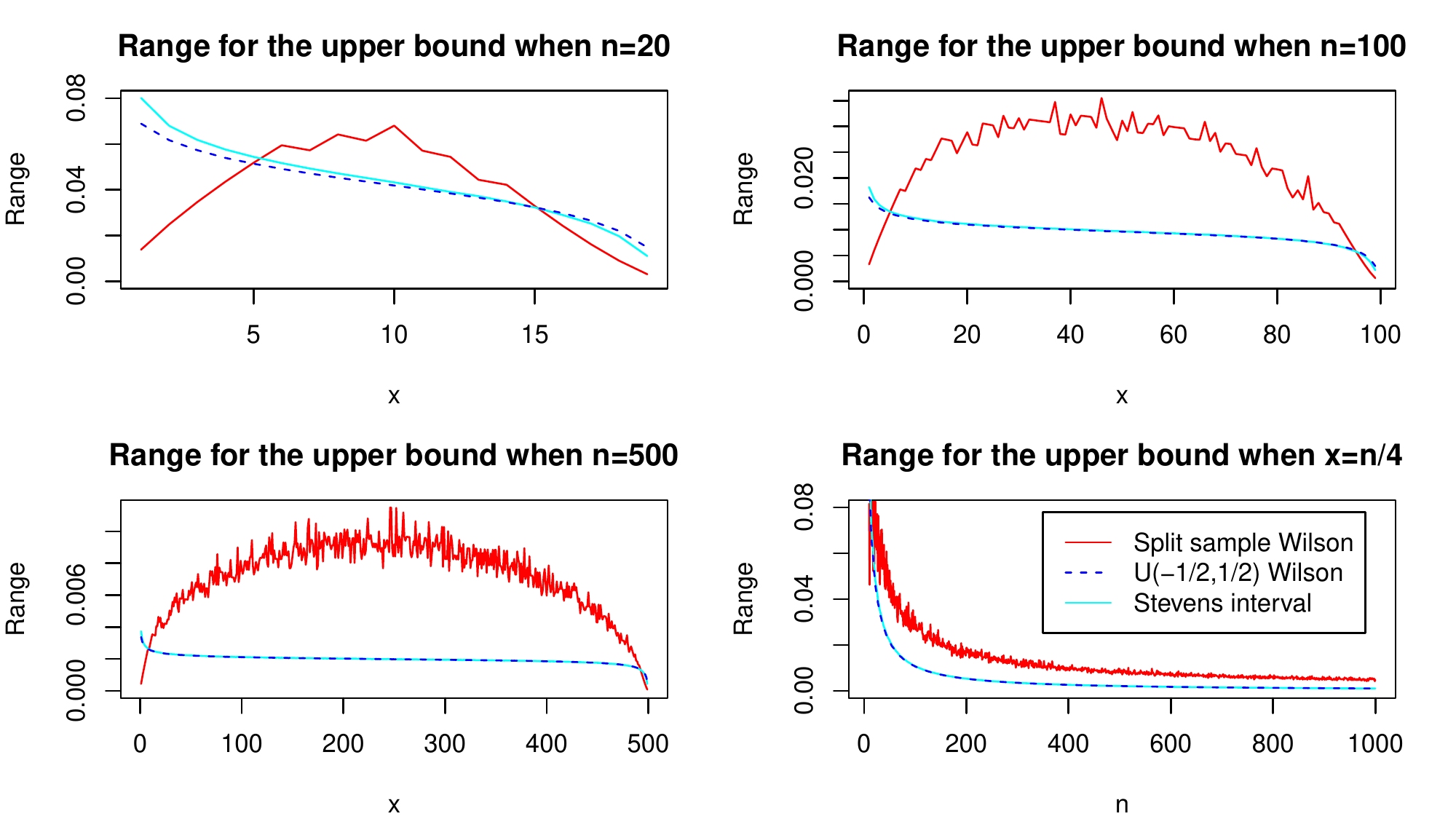}
 \end{center}
\end{figure}

The $U(-1/2,1/2)$ Wilson and Stevens intervals are similarly affected by the randomization for small $n$: when $n=10$ and $X=2$ the conditional expected length of the Stevens interval is 0.48 and the range of the upper bound is $0.111$ or 23 \% of the conditional expected length. However, the impact of the randomization on these interval decreases quicker than it does for the split sample Wilson interval. When $n=200$ and $X=40$ the conditional expected length of the Stevens interval is 0.11. The range of the upper bound is 0.005, which is no more than 4.5 \% of the conditional expected length.

Figure \ref{fig3} shows the range of the upper bounds of the three intervals for different combinations of $X$ and $n$. The intervals are about as bad for $n=20$, but the $U(-1/2,1/2)$ Wilson and Stevens intervals are much less affected by the randomization for $n=100$ and $n=500$. Because of the equivariance of the intervals, the figures for the lower bound are identical if $X$ is replaced by $n-X$.

\section{Data-randomized intervals}\label{datarandomized}

\subsection{Modifying externally randomized intervals}
As pointed out by \citet{hall14}, if the result of each of the $n$ independent trials are recorded, one can put the first $n_1$ observations in the first subsample and the other $n_2$ observations in the second subsample, so that the randomness only comes from the data itself. This means that there is no ambiguity about the interval, so that there is no need to discuss the range of the upper bound and that two different statisticians always will obtain the same confidence interval when they apply the procedure to the same data.

Alternatives to the split sample interval are available also if the sequence of observations has been recorded. It is possible to let the randomization of the Stevens interval be determined by $X_1,\ldots,X_n$, for instance by letting the sequence determine the (now approximately) $\mbox{U}(0,1)$ random variable $\nu_1$. An example of this is given by the following algorithm. It generalizes the data-randomized Korn interval \citep{korn87}, for which the p-value of a rank test of the hypothesis that the 1's are uniformly distributed in the sequence is used to determine $U_k$. It is our hope that the below description of the procedure is somewhat more straightforward.
\noindent\begin{center}\framebox[\width][c]{
\begin{minipage}{\textwidth}
\subsubsection*{Algorithm: Data-randomization of the Stevens interval.}
\begin{enumerate}
\item Compute all ${n}\choose{X}$ distinct permutations of $X_1,\ldots,X_n$.
\item Enumerate the permutations $k=1,\ldots,{{n}\choose{X}}$ and associate with each permutation a number $U_k=k/ {{n}\choose{X}}$.\\\small{\emph{Remark.} The enumeration can for instance be done by interpreting $X_1,\ldots,X_n$ as the base 2 decimal expansion of a number between 0 and 1, sorting the sequences according to this decimal expansion. The smallest $U_k$ is then assigned to the sequence corresponding to the smallest number, the second smallest $U_k$ to the sequence corresponding to the second smallest number, and so on.}
\item For the $k^*$ corresponding to the observed sequence $X_1,\ldots,X_n$, compute the Stevens interval with $\nu_1=U_{k^*}$ and $\nu_2=1-U_{k^*}$.
\end{enumerate}
\end{minipage}
}\end{center}
The algorithm can be applied to the $\mbox{U}(-1/2,1/2)$ Wilson interval as well, with $Y=U_k-1/2$.

The function used to associate a number $U_k$ with each permutation is arbitrary. However, as long as this function is given explicitly, no ambiguity will remain when the above algorithm is used. We also note that the split sample function proposed by \citet{hall14} is equally arbitrary, as is the choice of $n_1$ and $n_2$.

Data-randomized Stevens intervals can be constructed for parameters of other stochastically increasing lattice distributions, including the Poisson and negative binomial distributions, in an analogue manner. Similarly, the data-randomized $\mbox{U}(-1/2,1/2)$ procedure can be used to lessen the coverage fluctuations of other confidence intervals and for other distributions.

\subsection{Comparison of data-randomized intervals}\label{comp2}
When $n=20$ the split sample method smoothens the binomial distribution by replacing its 21 possible values by 119 possible values. When the above algorithm is used to produce the Korn interval, $\sum_{k=0}^{20}\binom{20}{k}=1,048,576$ possible values are used instead, smoothing the distribution even further, and reducing the coverage oscillations to a much greater extent.

\begin{figure}
\begin{center}
 \caption{Coverage and expected length of three data-randomized confidence intervals, compared to the non-randomized Wilson interval when $n=20$. The expected length of the Wilson interval equals that of the split sample Wilson interval and the data-randomized $\mbox{U}(-1/2,1/2)$ interval.}\label{fig4}
   \includegraphics[width=\textwidth]{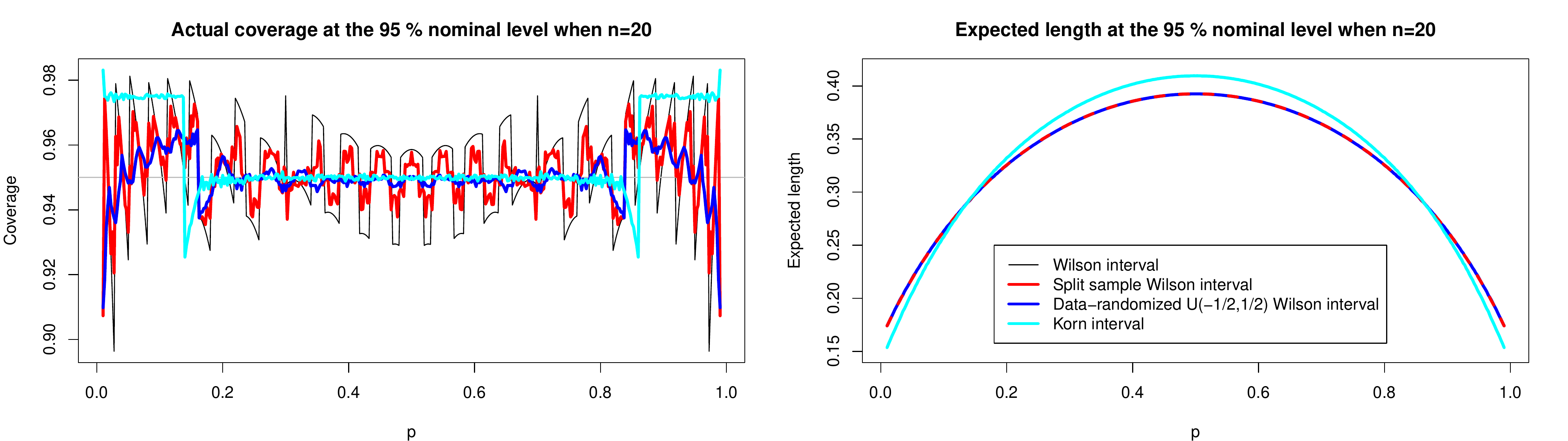}
 \end{center}
\end{figure}

The split sample Wilson interval is compared to the Korn and $\mbox{U}(-1/2,1/2)$ Wilson intervals in Figure \ref{fig4}. When $n=20$, the Korn interval has near-perfect coverage for $p\in(0.2,0.8)$, but like the externally randomized Stevens interval it is slightly wider than the competing intervals in this region of the parameter space. This difference in expected length quickly becomes smaller as $n$ increases. The data-randomized $\mbox{U}(-1/2,1/2)$ Wilson interval has smaller coverage fluctuations than the split sample Wilson interval, but is no wider.

\section{Discussion}\label{discussion}
In the split sample method random noise with a discrete distribution is added to the data to reduce the coverage fluctuations of a confidence interval. We have shown that it is better to use a continuous distribution such as the $\mbox{U}(-1/2,1/2)$ distribution for the random noise, as this lowers the coverage fluctuations even further and reduces the impact of the randomization on the interval bounds. This kind of randomization can however not be used to remove the coverage fluctuations completely. In contrast, the Stevens interval offers exact coverage. Moreover, for larger $n$ the randomization has a smaller impact on the bounds of the Stevens interval than it has on the split sample Wilson interval. If a randomized confidence interval for the binomial proportion $p$ is to be used, we therefore recommend the Stevens interval over split sample intervals.

The main objection against randomized confidence intervals is that different statisticians may produce different intervals when applying the same method to the same data. \citet{hall14} solve this problem by putting the first $n_1$ observations in the first subsample and the remaining $n_2$ observations in the second subsample, using so-called data-randomization. We compared several data-randomized intervals in Section \ref{comp2} and found that the split sample Wilson interval had the worst performance of the lot. Neither of these intervals manages to remove the coverage fluctuations completely.

When using the data-randomization strategy we have to condition our inference on an auxiliary random variable -- the order of the observations -- which has no relation to the parameter of interest. We then lose the rather natural property of invariance under permutations of the sample. Consequently, a statistician that reads the sequence from the left to the right may end up with a different interval than a statistician that reads the sequence from the right to the left. Care should be taken when using data-randomization: in most binomial experiments it is arguably neither reasonable nor desirable that the order in which the observations were obtained affects the inference. 

Objections against randomized confidence intervals are sometimes met by the argument that randomization already is widely used in virtually all areas of statistics: bootstrap methods, MCMC, tests based on simulated null distributions and random subspace methods (e.g. \citet{thu3}) all rely on randomization. An important difference is however that for any $n$ the impact of the randomization on these methods can be made arbitrarily small by allowing more time for the computations. This is not possible for randomized confidence intervals for parameters of lattice distributions, where for each $n$ there exists a positive lower bound for the impact of the randomization. In this context we also mention \citet{gm1}, who proposed a completely different approach to dealing with the ambiguity of randomized confidence intervals, based on fuzzy set theory.

In our study we applied the split sample method to the Wilson interval for a binomial proportion. This interval has excellent performance, with competitive coverage and length properties, and is often recommended for general use \citep{bcd1,newcombe12}. We note however that it can be more or less uniformly improved upon by using a level-adjusted Clopper--Pearson interval \citep{thu1} instead; we opted to use the Wilson interval in our study simply because it is less computationally complex.

\subsubsection*{Acknowledgement}
The author wishes to thank Silvelyn Zwanzig for helpful discussions.

\pagestyle{plain}

~\\[3mm]
Contact information:\\
M{\aa}ns Thulin, Department of Mathematics, Uppsala University, Box 480, 751 06 Uppsala, Sweden. E-mail: thulin@math.uu.se
\end{document}